\documentclass[conference]{IEEEtran}
\usepackage{cite}
\usepackage{amsmath,amssymb,amsfonts}
\usepackage{algorithmic}
\usepackage{graphicx}
\usepackage{textcomp}
\usepackage{xcolor}
\def\BibTeX{{\rm B\kern-.05em{\sc i\kern-.025em b}\kern-.08em
    T\kern-.1667em\lower.7ex\hbox{E}\kern-.125emX}}
    
\newcommand\blfootnote[1]{%
  \begingroup
  \renewcommand\thefootnote{}\footnote{#1}%
  \addtocounter{footnote}{-1}%
  \endgroup
}

\begin{document}

\title{Learning Modulation Design for SWIPT with Nonlinear Energy Harvester: Large and Small Signal Power Regimes}

\author{Morteza Varasteh$^\dagger$, Jakob Hoydis$^*$ and Bruno Clerckx$^\dagger$\\
$^\dagger$ Department of Electrical and Electronic Engineering, Imperial College London, UK.\\
$^*$ Nokia Bell Labs, Nozay, France\\
\{m.varasteh12; b.clerckx\}@imperial.ac.uk, jakob.hoydis@nokia-bell-labs.com.\vspace{-3mm}}

\maketitle

\begin{abstract}
Nonlinear \textit{energy harvesters} (EH) behave differently depending on the range of their input power. In the literature, different models have been proposed mainly for relatively small and large input power regimes of an EH. Due to the complexity of the proposed nonlinear models, obtaining analytical optimal or well performing signal designs have been extremely challenging. Relying on the proposed models in the literature, the learning problem of modulation design for \textit{simultaneous wireless information-power transfer} (SWIPT) over a point-to-point link is studied. Joint optimization of the transmitter and the receiver is implemented using \textit{neural network} (NN)-based autoencoders. The results reveal that for relatively small channel input powers, as the power demand increases at the receiver, one of the symbols is shot away from the origin while the remaining symbols approach zero amplitude. In the very extreme case of merely receiver power demand, the modulations are in the form of On-Off keying signalling with a low probability of the On signal. On the other side, for relatively large channel input powers, it is observed that as the receiver power demand increases, a number of symbols approach zero amplitude, whereas the others (more than one symbol) get equally high amplitudes but with different phases. In the extreme scenario of merely receiver power demand, the modulation resembles multiple On-Off keying signalling with different phases.\blfootnote{This work has been partially supported by the EPSRC of the UK under grant EP/P003885/1.}
\end{abstract}

\vspace{-2mm}
\section{Introduction}\label{Sec_Intro}

In order to design efficient \textit{simultaneous wireless information and power transfer} (SWIPT) architectures, it is crucial to model the \textit{energy harvester} (EH) with a high level of accuracy. The EH consists of a rectenna, which is composed of an antenna followed by a rectifier. The rectifier is used to convert the \textit{radio frequency} (RF) power into \textit{direct current} (DC) current in order to charge low power devices. Although most of the results in the literature adopt a linear characteristic function for the rectifier, in practice, due to the presence of a diode in the rectifier, the output of the EH is a nonlinear function of its input \cite{Clerckx_Bayguzina_2016, Boaventura_Collado_Carvalho}. Due to the nonlinearity of the diode characteristic function, the RF-to-DC conversion efficiency of the EH is highly dependent on the power as well as the shape of the waveform \cite{Clerckx_Bayguzina_2016,Boaventura_Collado_Carvalho,Clerckx_Bayguzina_2017}. Observations based on experimental results reveal that signals with high \textit{peak-to-average power ratio} (PAPR) result in high delivered DC power compared to other signals \cite{Boaventura_Collado_Carvalho}. In the literature, so far, depending on the application and available resources, two different models for a nonlinear EH are proposed. The first model is based on Taylor expansion of the diode characteristic function and is introduced in \cite{Clerckx_Bayguzina_2016}. It is shown that the harvested power is a function of summation of even moments of the received RF signal, which can be approximated with an acceptable level of accuracy by truncating it to the second and fourth moments. This model is appropriate for applications where the EH operates in a range that within which the diode is unlikely to break down. The other model, which is a function of the RF power of the received signal, is introduced in \cite{Boshkovska}. In this model, the rectenna's input/output power relationship is expressed in terms of a sigmoidal function. This model is appropriate in applications where there is little guarantee for the EH to operate below the diode breakdown edge.

In SWIPT systems, the goal is to maximize the DC power as well as the information rate, which is commonly referred to maximizing the achievable \textit{rate-power} (RP) region. Unlike most of the SWIPT systems with the linear model assumption for EH, for SWIPT systems with nonlinear EH, there exists a tradeoff between the rate and delivered power \cite{Clerckx_Zhang_Schober_Wing_Kim_Vincent}. Due to the presence of nonlinear components in EH, obtaining the exact optimal tradeoff analytically has so far been unsuccessful. However, after making some simplifying assumptions, some interesting results have been derived in \cite{Varasteh_Rassouli_Clerckx_FS_arxiv,Clerckx_2016,Varasteh_Rassouli_Clerckx_ITW_2017,Morsi_Jamali,Varasteh_Rassouli_Clerckx_arxiv}. In particular, in multicarrier transmission, it is shown in \cite{Varasteh_Rassouli_Clerckx_FS_arxiv,Clerckx_2016} that nonzero mean Gaussian input distributions lead to an enlarged RP region compared to \textit{circularly symmetric complex Gaussian} (CSCG) input distributions. In single carrier transmissions over \textit{additive white Gaussian noise} (AWGN) channel, in  \cite{Varasteh_Rassouli_Clerckx_arxiv,Morsi_Jamali}, it is shown that (under nonlinearity assumption for the EH) for circular symmetric inputs, the capacity achieving input distribution is discrete in amplitude with a finite number of mass-points and with a uniformly distributed independent phase. This is in contrast to the linear model assumption of the EH, where there is no tradeoff between the information and power (i.e., from system design perspective the two goals are aligned), and the optimal inputs are Gaussian \cite{Clerckx_Zhang_Schober_Wing_Kim_Vincent}.

While designing SWIPT signals and systems (under nonlinear assumptions for the EH) using analytical tools seems extremely cumbersome, learning (L)-based methods reveal a promising alternative to tackle the aforementioned problems. In fact, L-based methods, and particularly, autoencoders have recently shown remarkable results in communications, achieving or even surpassing the performance of state-of-the-art algorithms \cite{OShea_Hoydis_2017}. In particular, the similarities between the autoencoder architecture and the digital communications systems have motivated significant research efforts in the direction of modelling end-to-end communication systems using the autoencoder architectures \cite{OShea_Hoydis_2017}. Some examples include decoder design for existing channel codes \cite{Nachmani_etall}, blind channel equalization \cite{Caciularu_Burshtein}, learning physical layer signal representation for SISO \cite{OShea_Hoydis_2017} and, OFDM systems \cite{Felix_Cammerer_Dorner}.

\begin{figure}
\begin{centering}
\includegraphics[scale=0.39]{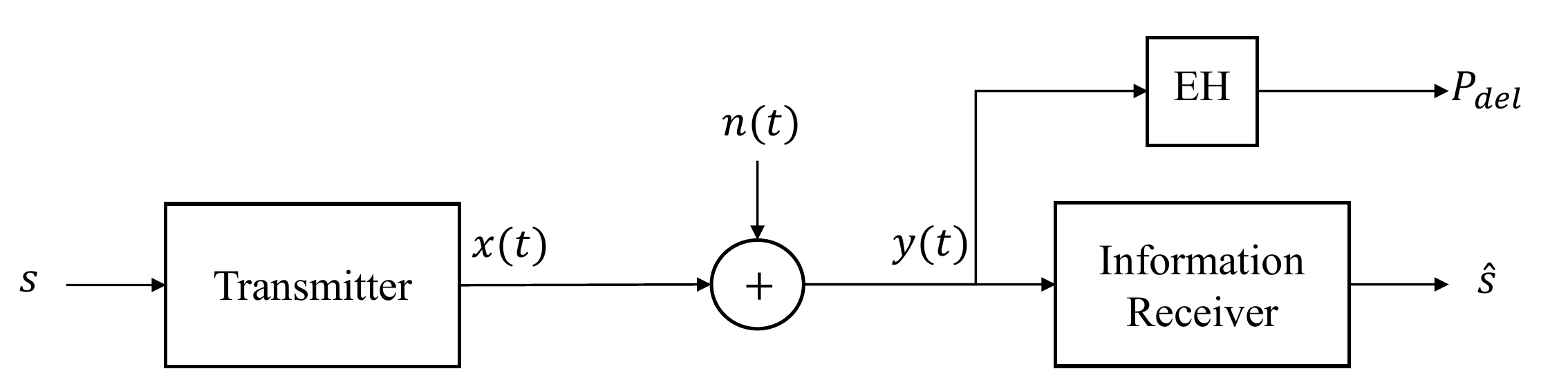}
\caption{Point-to-point SWIPT architecture}\label{Fig_2}
\par\end{centering}
\vspace{-3mm}
\end{figure}

In this work, we consider signal modulation design for a point-to-point SWIPT over a noisy channel in two different design regimes, namely, small and large EH input power. In particular, we consider the SWIPT system as an autoencoder structure, where the transmitter and the receiver are considered as multi-layer \textit{neural networks} (NN). The numerical optimizations reveal that in the small power regime, by increasing the power demand at the receiver, all but one of the channel input symbols (denoted as information symbols) converge towards zero amplitude, whereas the other symbol (denoted as power symbol) gets away from zero. Also, the power symbol is always along either the real or imaginary axis. This observation is inline with the result in \cite{Varasteh_Rassouli_Clerckx_ITW_2017}, where it is shown that as the power demand at the receiver increases, the transmitter allocates more power to either real or imaginary axis. It is also inline with the On-Off keying signalling interpretation of \cite{Varasteh_Rassouli_Clerckx_arxiv}. Similar results are reported in \cite{Varasteh_Piovano_Clerckx} for another model (proposed in \cite{Vedady_Zeng}) of the EH in the low input power regime. For the large input power regime, it is observed that more than one symbol get away from zero by increasing the receiver power demand, whereas the other symbols approach towards zero amplitude. This resembles a scenario where the transmitter favours multiple On-Off keying signalling with smaller amplitudes rather than one with higher amplitude.

The rest of the paper is organized as follows. In Sec. \ref{Sec:sys_Model}, we introduce the studied problem. In Sec. \ref{Sec:EH}, we review the two EH models proposed in the literature that we use in this paper. Implementation of the studied system using NN-based autoencoder is provided in Sec. \ref{Sec:Implemen}. Numerical results are provided in Sec. \ref{Sec:Numeric} following the conclusion in Sec. \ref{Sec:conclu}.


\section{System Model}\label{Sec:sys_Model}
A point-to-point SWIPT architecture is shown in Fig. \ref{Fig_2}. The receiver harvests the power (denoted by $P_{\text{del}}$) of the received signal and decodes the information, jointly. The baseband information bearing pulse modulated signal is represented as $x(t) = \sum_{k=-\infty}^{\infty}\pmb{x}g(t-kT_s)$, where $g(t)$ and $\pmb{x}$ are the pulse waveform and the complex information-power symbol, respectively. The complex information-power symbols $\pmb{x}$ are under an average power constraint, i.e., $\mathbb{E}[|$\pmb{x}$|^2]\leq P_a$. The received signal in the baseband (from which the transmitted message is estimated) is $y(t) = x(t)+n(t)$, where $n(t)$ is the baseband complex-valued noise. The EH is fed with the received RF signal, i.e., $y_{\text{RF}}(t)=\sqrt{2}\text{Re}\{y(t)e^{j2\pi f_c t}\}$, where $f_c$ is the carrier frequency. In the following section, we review the two EH models studied in this paper.

\begin{figure}
\begin{centering}
\includegraphics[scale=0.42]{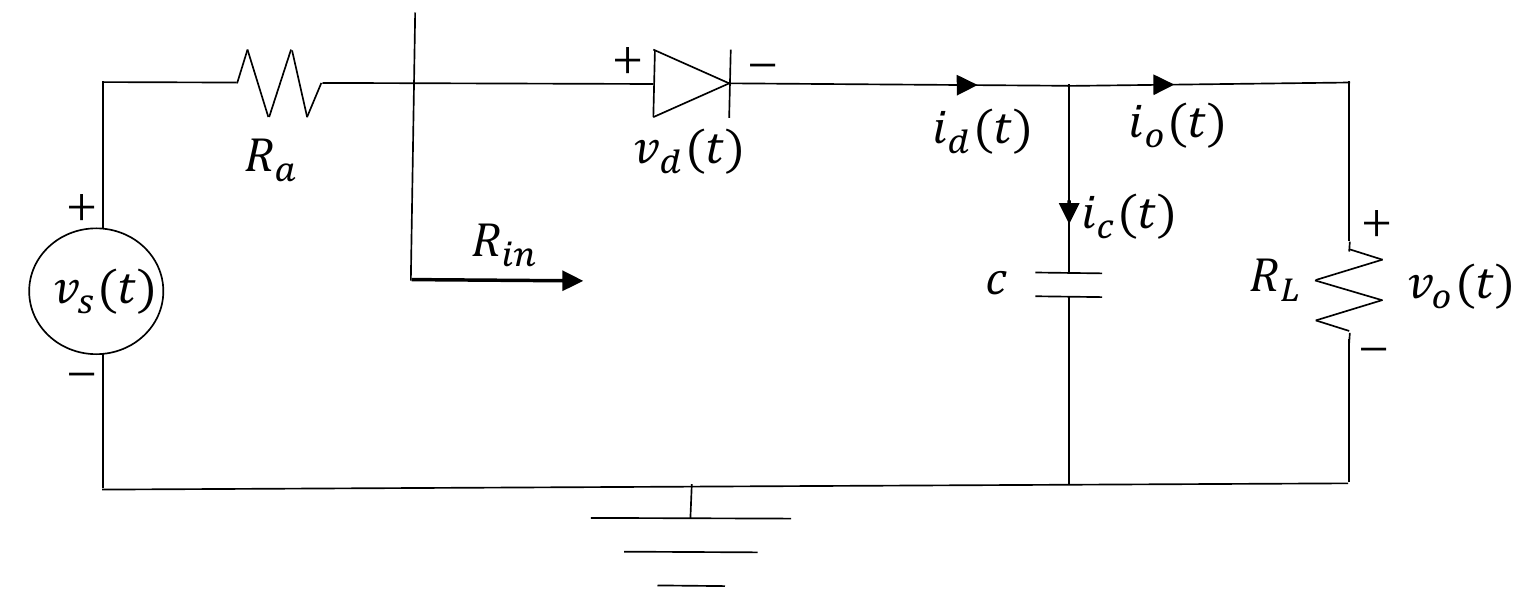}
\caption{The model for the rectenna circuit of a practical EH}\label{Fig_1}
\par\end{centering}
\vspace{-3mm}
\end{figure}


\section{EH Models}\label{Sec:EH}
A simple structure of a rectenna is illustrated in Fig. \ref{Fig_1}. The received RF signal $y_{\text{RF}}(t)$ is converted at the rectifier's output into a DC signal across a load resistance $R_L$. In the literature (depending on the application and available resources), different analytic models of the EH have been proposed accounting for the nonlinearity. These models can be mainly categorized into two groups, namely, small and large EH input power regimes. The main focus of the rectifier in the small power regime (about $-30$dBm to $-10$dBm input power) is to operate in the square law and transition zones \cite{Boaventura_Collado_Carvalho,Clerckx_Zhang_Schober_Wing_Kim_Vincent}    , which also captures the \textit{threshold effect}, i.e. the power level below which the rectifier turns off. However, for the models suitable for large power regime (above $-10$dBm input power), the main focus is to capture the \textit{saturation effect}, i.e., the power level above which the rectifier's output DC power gain remains constant.\footnote{Operating diodes in the breakdown region is not the purpose of a rectifier and should be avoided as much as possible \cite{Boaventura_Collado_Carvalho,Clerckx_2016,Clerckx_Zhang_Schober_Wing_Kim_Vincent}} Next, we review the two existing models in the literature in Secs \ref{Sec:Model_1}, \ref{Sec:Model_2} used for small and large input power regimes of an EH, respectively.


\subsection{Model $A$ (Small Power Regime)}\label{Sec:Model_1}
This model is based on the Taylor expansion of the diode characteristic function and is introduced in \cite{Clerckx_Bayguzina_2016}. It is shown that the delivered power is a function of summation of even moments of the received RF signal, which can be approximated with a high level of accuracy by truncating it to the second and fourth moments. Accordingly, the delivered power\footnote{According to \cite{Clerckx_2016}, rectenna's output is in the form of current with unit Ampere. However, since power is proportional to current, with abuse of notation, we refer to the term in (\ref{Eq_1}) as power.}, denoted by $P_{\text{del}}$, is modelled as
\begin{align}\label{Eq_1}
P_{\text{del}}=\mathbb{E}[\mathcal{E}[k_2y_{\text{RF}}(t)^2 + k_4 y_{\text{RF}}(t)^4]]
\end{align}
where $k_2$ and $k_4$ are constants and $\mathbb{E}[\cdot],~\mathcal{E}[\cdot]$ are the expectation operator over the randomness of the signal and the averaging operator over time, respectively. Note that the delivered power model introduced in (\ref{Eq_1}) is in the RF domain. From a communications system design point of view, it is preferable to have a baseband equivalent representation of the system. In \cite[Proposition 3]{Varasteh_Rassouli_Clerckx_arxiv}, it is shown that the baseband equivalent of (\ref{Eq_1}) for \textit{independent and identically distributed} (iid) inputs is obtained as
\begin{align}\label{Eq_2}
P_{\text{del}}=\alpha (Q+\tilde{Q})+\beta P+\gamma
\end{align}
where the parameters $\alpha,~\beta$, and $\gamma$ are constants (for more detail see \cite{Varasteh_Rassouli_Clerckx_arxiv}) and $\tilde{Q}$ is given by
\begin{align}\nonumber
\tilde{Q}&=\frac{1}{3}\big(Q_{r}+Q_{i}+2(\mu_{r}T_{r}+\mu_{i}T_{i})\\
&+6P_{r}P_{i}+6P_{r}(P_{r}-\mu_{r}^{2})+6P_{i}(P_{i}-\mu_{i}^{2})\big)
\end{align}
with $Q=\mathbb{E}[|\pmb{x}|^4]$, $T=\mathbb{E}[|\pmb{x}|^3]$, $P=\mathbb{E}[|\pmb{x}|^2]$, $\mu=\mathbb{E}[\pmb{x}]$. Similarly, $Q_r=\mathbb{E}[\pmb{x}_r^4]$, $T_r=\mathbb{E}[\pmb{x}_r^3]$, $P_r=\mathbb{E}[\pmb{x}_r^2]$, $\mu_r=\mathbb{E}[\pmb{x}_r]$ and $Q_i=\mathbb{E}[\pmb{x}_i^4]$, $T_i=\mathbb{E}[\pmb{x}_i^3]$, $P_i=\mathbb{E}[\pmb{x}_i^2]$, $\mu_i=\mathbb{E}[\pmb{x}_i]$, where $\pmb{x}_r$, $\pmb{x}_i$ are the real and imaginary components of $\pmb{x}$, respectively.


\subsection{Model $B$ (Large Power Regime)}\label{Sec:Model_2}
Although for low input power of the EH, the RF energy conversion efficiency improves as the input power increases, there are diminishing returns and limitations on the maximum possible harvested power whenever the same rectifier is used for the low and high power regime \cite{Boaventura_Collado_Carvalho,Valenta_Durgin}.\footnote{Adapting the rectifier as the input power level increases would avoid the saturation problem \cite{Boaventura_Collado_Carvalho}.} \cite{Valenta_Durgin}. Accordingly, in \cite{Boshkovska}, a nonlinear model based on the sigmoidal function is proposed that captures this practical limitation. The harvested power at the output of the EH is modelled as
\begin{align}\label{Eq_3}
P_{\text{del}}=\frac{\Psi-L_s\Omega}{1-\Omega}
\end{align}
where $\Omega=1/(1+ \exp(a\cdot b))$ and $\Psi=L_s/(1+\exp(-a\cdot(P_{\text{in}}-b)))$ with $P_{\text{in}}$ denoted as the EH input power. The parameter $\Omega$ is to ensure the zero-input/zero-output response for the EH, and the parameter $L_s$ denotes the maximum harvested power when the EH is saturated. In this paper, we adopt the values $L_s=0.02,~a=6400$, and $b=0.003$ as reported in \cite{Boshkovska}.

\begin{figure}
\begin{centering}
\includegraphics[scale=0.49]{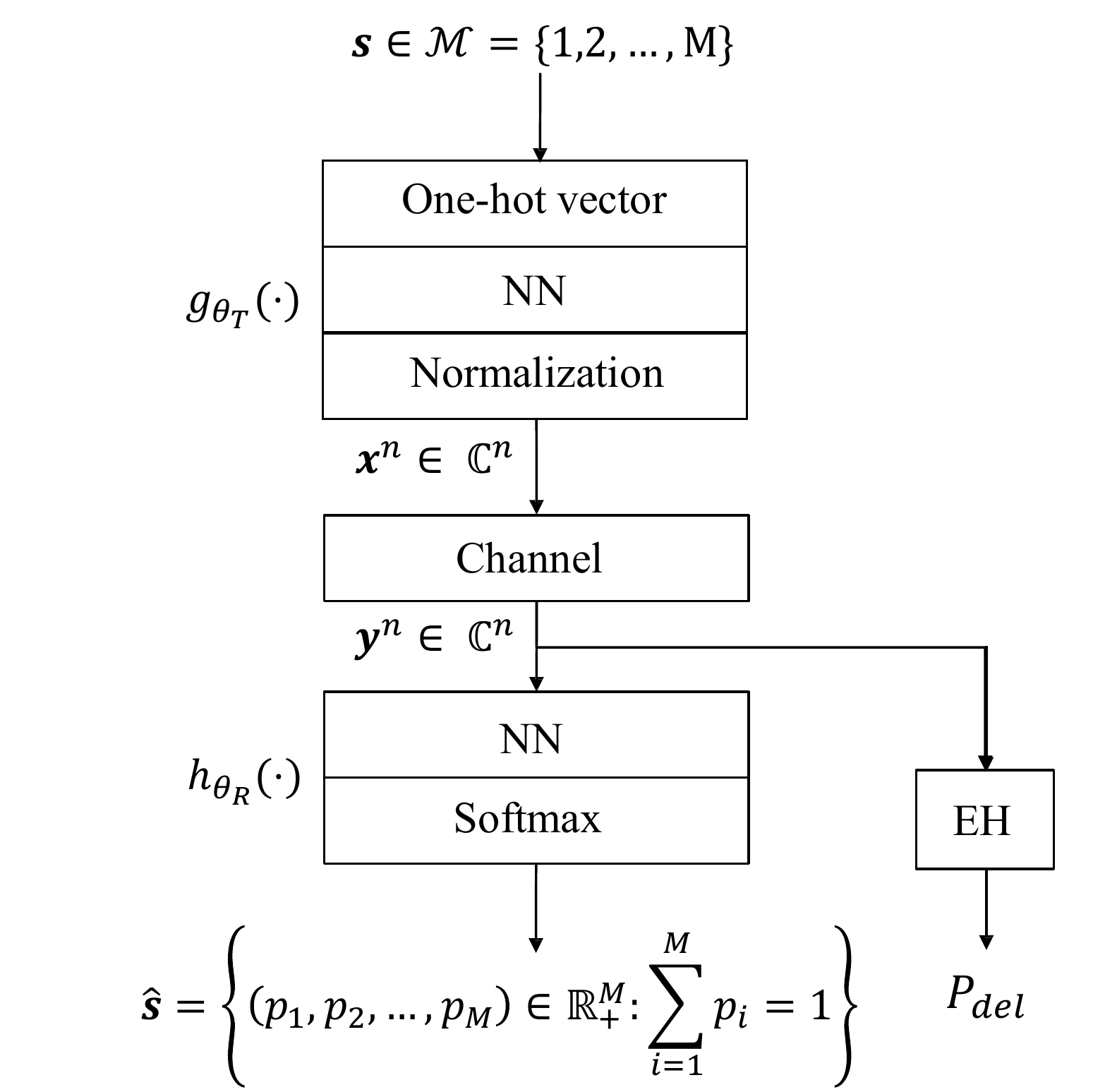}
\caption{Illustration of the autoencoder structure}\label{Fig_3}
\par\end{centering}
\vspace{-4mm}
\end{figure}


\section{Implementation}\label{Sec:Implemen}
We consider the SWIPT system in Fig. \ref{Fig_2} as an NN-based autoencoder, where both the transmitter and receiver are implemented as two fully connected NNs in order to perform the modulation and demodulation processes, respectively. The transmitter communicates one of $M$ possible messages $s \in \mathcal{M} = \{1,2,...,M\}$, where $\mathcal{M}$ denotes the message alphabet set. First the message $s \in \mathcal{M}$ is converted into a one-hot vector\footnote{An $M$-dimensional vector of all zeros except one in $s^{\text{th}}$ position} denoted by $\mathbf{s}$. The vector $\mathbf{s}$ is then converted into a complex codeword $\pmb{x}^{n}\in \mathcal{X}^n$ consisting of $n$ complex information-power symbols (here we consider $n=1$). The mapping from the set of messages $\mathcal{M}$ to the transmitted signal space $\mathcal{X}^n$ is denoted by $g_{\theta_T}(\cdot): \mathcal{M} \rightarrow \mathbb{C}^n$, where $\theta_T$ refers to the set of transmitter parameters, related to the weights and biases across the layers of the NN. In order to satisfy the average power constraint at the transmitter, a power normalization is performed as the last layer of the transmitter. The modulated signal $\pmb{x}^n$ is corrupted by the channel noise (here we consider AWGN).

The receiver aims both at detecting the transmitted symbol $s$ as well as harvesting the power $P_{\text{del}}$ of the received signal denoted as $\pmb{y}^n$, where $\pmb{y}^n$ is the n-length samples of the signal $y(t)$ taken with frequency $1/T$ Hz. The detection is performed by mapping the received noisy codeword $\pmb{y}^n$ to an $M$-dimensional probability vector denoted by $\hat{\mathbf{s}}$ (and outputting the detected message by obtaining the index corresponding to the maximum probability) through a parametric function denoted by $h_{\theta_R}(\cdot): \mathbb{C}^n \rightarrow \mathcal{M}$ and implemented as a fully-connected NN. $\theta_R$ refers to the set of receiver parameters in terms the weights and biases across the decoder module.

\begin{figure}
\begin{centering}
  \includegraphics[scale=0.41]{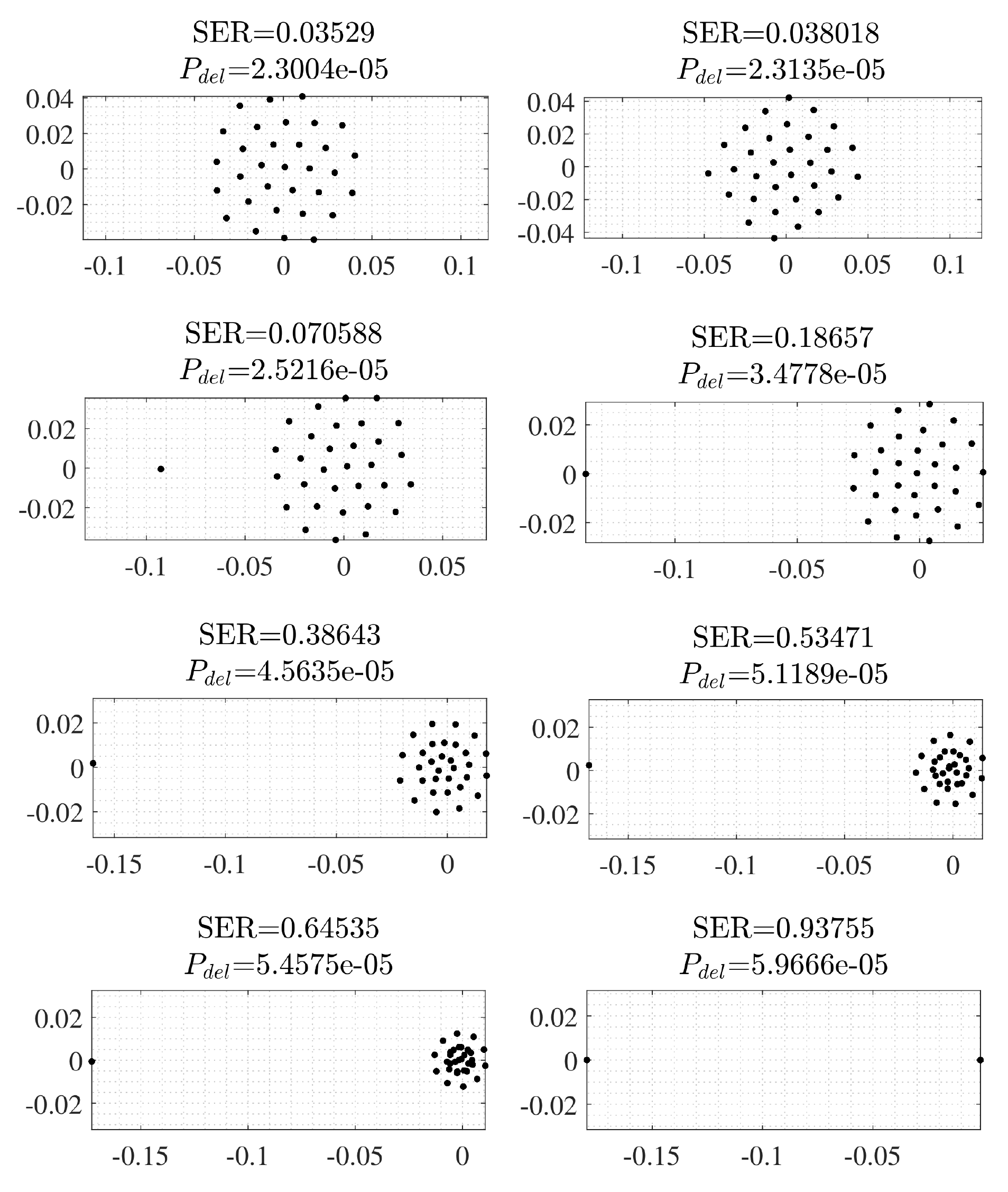}
  \caption{Modulations for different receiver information and power demands (for model A) with $M=32$ messages, average power constraint $P_a=0.001$, and $\text{SNR}=50$.}\label{Fig_4}
  \par\end{centering}
  \vspace{-3mm}
\end{figure}

The delivered power $P_{\text{del}}$ at the receiver is modelled as in (\ref{Eq_2}) and (\ref{Eq_3}) for small and large EH input power regimes, respectively. Since for power delivery purposes, the received RF signal is directly fed into the EH, the signal is not processed through the NN. We model the information loss as the cross entropy function between the transmitted one-hot vector $\mathbf{s}$, and the output probability vector $\mathbf{\hat{s}}$ at the receiver, i.e.,  $\mathcal{L}(\mathbf{s}, \mathbf{\hat{s}}) = -\sum_{i=1}^M {\mathbf{s}_i \log \hat{\mathbf{s}}_i}$, where $\mathbf{s}_i$ and $\hat{\mathbf{s}}_i$ indicate the $i^{\text{th}}$ entry of the vectors $\mathbf{s}$ and $\hat{\mathbf{s}}$, respectively. Accordingly, the cost function used in order to optimize the system is given by
\begin{align} \label{Eq_4}
L(\theta_T,\theta_R) = \frac{1}{|\mathcal{B}^m|}\sum_{k\in \mathcal{B}^m} \mathcal{L}(\mathbf{s}^{(k)}, \mathbf{\hat{s}}^{(k)}) +\frac{\lambda}{P_{\text{del}}}
\end{align}
where $\mathcal{B}^m$ is a randomly drawn minibatch of training data, which is assumed to be generated iid with a uniform distribution over the message set. Note that different values of the parameter $\lambda\geq 0$ in (\ref{Eq_4}) can be associated to different information-power demands at the receiver. In our implementation, we have used stochastic gradient descent with the Adam optimizer. All code was written in Python using Tensorflow. We note that it is crucial for the model considered for the EH to be differentiable. Alternatively, a real system can be optimized directly without modelling, using gradient estimation techniques (see, e.g., \cite{Hoydis_Aoudia}).


\begin{figure}
\begin{centering}
  \includegraphics[scale=0.5]{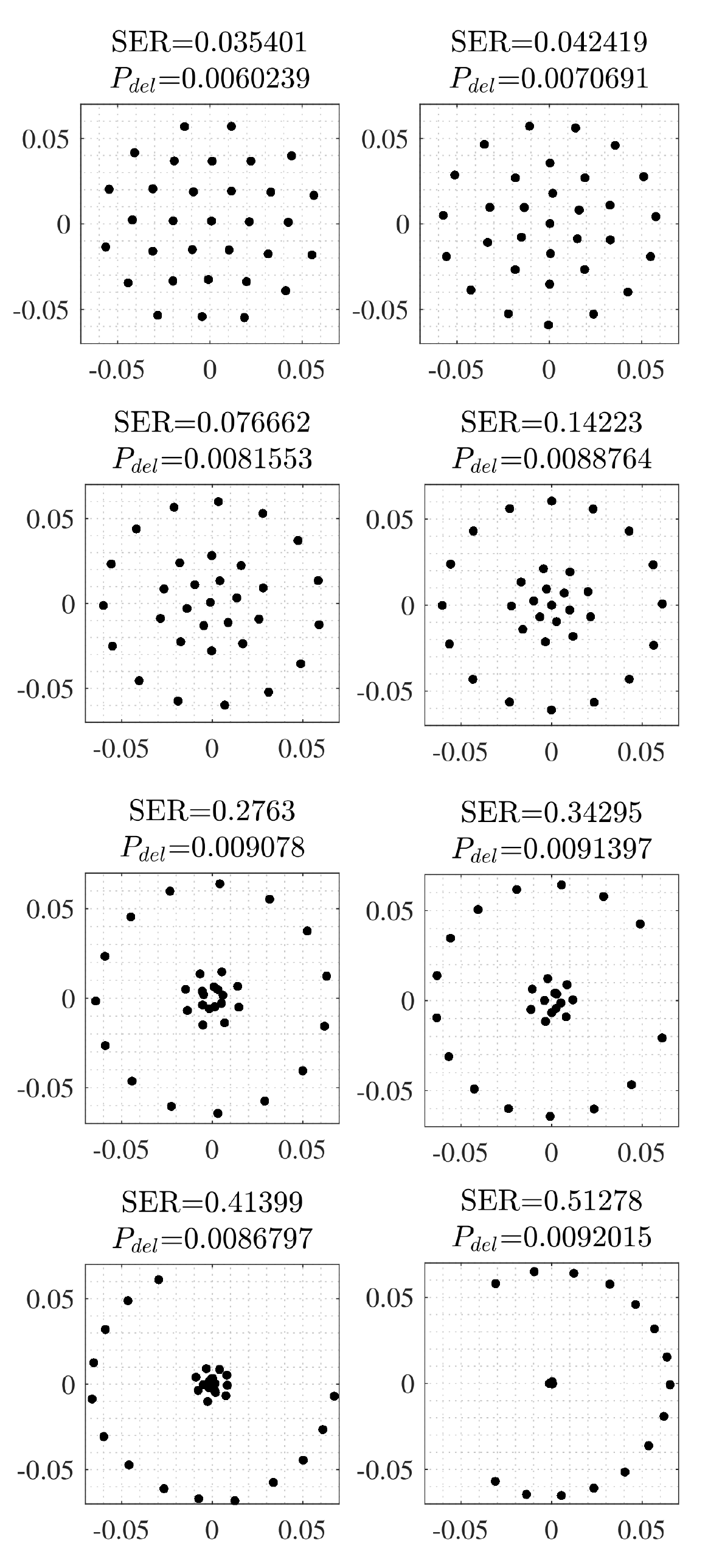}
  \caption{Modulations for different receiver information and power demands (for model B) with $M=32$ messages, average power constraint $P_a=0.002$, and $\text{SNR}=50$.}\label{Fig_5}
  \par\end{centering}
  \vspace{-3mm}
\end{figure}


\section{Numerical Results}\label{Sec:Numeric}

We consider a training set of $10^5\cdot M$ symbols (produced uniformly over the set of the transmitted messages) and a mini-batch size of $10^3\cdot M$. The optimization runs through $5000$ epochs. For the optimization, ADAM mini-batch gradient descent optimizer is considered with the starting learning rate of 0.01. Xavier and zero initialization are considered for the weights and the biases, respectively. For each value of $\lambda$, the model is run $100$ times and each time with a different initialization seed (so that the effect of local optima is alleviated). The final solution (well performing mapping for the transmitter and the receiver) is chosen as the one which minimizes the cost function in (\ref{Eq_4}) the most. For the final performance evaluation, we use a test set of size $5\times 10^6\cdot M$ (uniformly distributed over the message set). Since we do not consider coded modulation, we have $n=1$, i.e., a communication rate of $R = \log_2(M)$ bits per complex channel use.

The value of $\lambda$ is incrementally increased starting from zero (Note that $\lambda=0$ is equivalent to the receiver information-only demands). The termination of the optimization is performed as follows. We consider a certain threshold as the maximum allowable symbol error (SER) rate denoted as $\text{SER}_{\text{max}}$ (here we consider $\text{SER}_{\text{max}}=0.95$). We continue increasing $\lambda$ until the inequality SER $\leq\text{SER}_{\text{max}}$ is contradicted, where $\text{SER}$ is the symbol error rate corresponding to the learned system.

In Fig. \ref{Fig_4} and Fig. \ref{Fig_5}, the modulations for different information rate and power demands at the receiver are illustrated for model A and model B of a nonlinear EH, respectively. For model A, it is observed from Fig. \ref{Fig_4} that as $\lambda$ increases (in the direction of left to right and top to bottom, which is equivalent to increasing the power demand at the receiver), the modulation reshapes in a way that all but one of the symbols (denoted as information symbols) converge towards zero amplitude keeping their symmetric structure. The other symbol (denoted as power symbol), however, gets away from the origin. In the very extreme scenario, where the receiver only demands power, all but one of the symbols collapse on top of each other at the origin with the other symbol shot away from zero. This is indeed equivalent to having two symbols, one with a high probability around zero and the other with a small probability and high amplitude. This observation is similar to the results in \cite{Varasteh_Rassouli_Clerckx_arxiv}, where it is shown that for receiver merely power demands, the transmitted signal approaches to an On-Off keying signalling with a low probability of the On signal. It is also observed from the results in Fig. \ref{Fig_4} that the power symbol gets away along the real or imaginary axis (here all are along the real axis). This is due to the stronger effect of the fourth moment of the channel input (on harvested power via the EH) with respect to its second moment. This observation is also in line with the results in \cite{Varasteh_Rassouli_Clerckx_ITW_2017}, where it is shown that for a zero mean Gaussian (in general complex) input, if the receiver demand is only power, the power budget at the transmitter must be allocated to either real or imaginary subchannel.  From Fig. \ref{Fig_5}, under the assumption of model B for a nonlinear EH (large input power regime), it is observed that as $\lambda$ increases, the modulation reshapes in way that some of the symbols, get equally higher amplitudes with different phases, whereas the other symbols converge towards zero amplitude, until in the very extreme scenario of receiver merely power demand (the last plot in Fig. \ref{Fig_5}), they collapse on top of each other at zero. This can be explained as follows. From harvesting power perspective, (unlike model A) there is no benefit for the amplitude of the transmitted symbols being larger than a threshold (note that this threshold is determined by the saturation level of the EH, and based on our observations it is roughly equal to the square root of the saturation power level, i.e., $\sqrt{L_s}$). Accordingly, the optimization converges to a solution where multiple symbols (rather than one) get higher amplitudes with the remaining symbols collapsing on top of each other at zero amplitude (the last plot in Fig. \ref{Fig_5}). Indeed the transmitter chooses to have multiple On-Off keying signalling with relatively lower amplitudes of the On signals rather than having one with large amplitude (as in the small power regime).

We note that in the case of model A (as long as the EH is not saturated), the harvested power increases by the number of messages (under the same average power constraint $P_a$). This is due to the fact that better On-Off keying signalling (higher probability for the Off signal) is produced by higher number of constellation points. However, this is not the case for model B due to the saturation effect of the EH. Instead, for model B (even if the receiver demands only power), it is still possible to transmit information with an acceptable SER by neglecting the zero amplitude symbols.


\section{Conclusion}\label{Sec:conclu}
In this paper, we studied the problem of modulation design in a point-to-point SWIPT problem under two different nonlinear models for the EH. In particular, we considered two models proposed in the literature for small and large input power regimes of an EH. The modulations for the two models for different receiver information rate and power demands are obtained by jointly optimizing over the transmitter and the receiver that are considered as NN-based autoencoders. The results based on numerical optimization reveal that for the small input power regime of an EH, as the receiver power demand increases, the modulations converge to one On-Off keying signalling with a low probability of the On signal, whereas for large input power regime of an EH, the modulations converge to those with multiple On-Off keying signalling. Overall, we believe that learning-based approaches have huge potential for the optimization of SWIPT systems due to unwieldy EH models that are difficult to analyze analytically.
\vspace{-5mm}

\bibliographystyle{ieeetran}
\bibliography{ref_1}

\begin{thebibliography}{10}
\providecommand{\url}[1]{#1}
\csname url@samestyle\endcsname
\providecommand{\newblock}{\relax}
\providecommand{\bibinfo}[2]{#2}
\providecommand{\BIBentrySTDinterwordspacing}{\spaceskip=0pt\relax}
\providecommand{\BIBentryALTinterwordstretchfactor}{4}
\providecommand{\BIBentryALTinterwordspacing}{\spaceskip=\fontdimen2\font plus
\BIBentryALTinterwordstretchfactor\fontdimen3\font minus
  \fontdimen4\font\relax}
\providecommand{\BIBforeignlanguage}[2]{{%
\expandafter\ifx\csname l@#1\endcsname\relax
\typeout{** WARNING: IEEEtran.bst: No hyphenation pattern has been}%
\typeout{** loaded for the language `#1'. Using the pattern for}%
\typeout{** the default language instead.}%
\else
\language=\csname l@#1\endcsname
\fi
#2}}
\providecommand{\BIBdecl}{\relax}
\BIBdecl

\bibitem{Clerckx_Bayguzina_2016}
B.~Clerckx and E.~Bayguzina, ``Waveform design for wireless power transfer,''
  \emph{IEEE Trans. Signal Processing}, vol.~64, no.~23, pp. 6313--6328, Dec.
  2016.

\bibitem{Boaventura_Collado_Carvalho}
A.~Boaventura, A.~Collado, N.~B. Carvalho, and A.~Georgiadis, ``Optimum
  behavior: Wireless power transmission system design through behavioral models
  and efficient synthesis techniques,'' \emph{IEEE Microwave Magazine},
  vol.~14, no.~2, pp. 26--35, Mar. 2013.

\bibitem{Clerckx_Bayguzina_2017}
B.~Clerckx and E.~Bayguzina, ``Low-complexity adaptive multisine waveform
  design for wireless power transfer,'' \emph{IEEE Antennas and Wireless
  Propagation Letters}, vol.~16, pp. 2207--2210, 2017.

\bibitem{Boshkovska}
E.~Boshkovska, D.~W.~K. Ng, N.~Zlatanov, and R.~Schober, ``Practical non-linear
  energy harvesting model and resource allocation for {SWIPT} systems,''
  \emph{IEEE Commun. Letters}, vol.~19, no.~12, pp. 2082--2085, Dec. 2015.

\bibitem{Clerckx_Zhang_Schober_Wing_Kim_Vincent}
B.~Clerckx, R.~Zhang, R.~Schober, D.~W.~K. Ng, D.~I. Kim, and H.~V. Poor,
  ``Fundamentals of wireless information and power transfer: From {RF} energy
  harvester models to signal and system designs,'' \emph{IEEE Journal on
  Selected Areas in Communications}, vol.~37, no.~2, pp. 1--30, Feb. 2019.

\bibitem{Varasteh_Rassouli_Clerckx_FS_arxiv}
\BIBentryALTinterwordspacing
M.~Varasteh, B.~Rassouli, and B.~Clerckx, ``{SWIPT} signalling over
  frequency-selective channels with a nonlinear energy harvester: Non-zero mean
  and asymmetric inputs,'' \emph{CoRR}, vol. abs/1901.01740, 2019. [Online].
  Available: \url{http://arxiv.org/abs/1901.01740}
\BIBentrySTDinterwordspacing

\bibitem{Clerckx_2016}
B.~Clerckx, ``Wireless information and power transfer: Nonlinearity, waveform
  design, and rate-energy tradeoff,'' \emph{IEEE Trans. Signal Processing},
  vol.~66, no.~4, pp. 847--862, Feb. 2018.

\bibitem{Varasteh_Rassouli_Clerckx_ITW_2017}
M.~Varasteh, B.~Rassouli, and B.~Clerckx, ``Wireless information and power
  transfer over an {AWGN} channel: Nonlinearity and asymmetric gaussian
  signaling,'' in \emph{IEEE Inf. Theory Workshop (ITW)}, Nov. 2017, pp.
  181--185.

\bibitem{Morsi_Jamali}
R.~Morsi, V.~Jamali, D.~W.~K. Ng, and R.~Schober, ``On the capacity of {SWIPT}
  systems with a nonlinear energy harvesting circuit,'' in \emph{IEEE
  International Conference on Commun. (ICC)}, May 2018, pp. 1--7.

\bibitem{Varasteh_Rassouli_Clerckx_arxiv}
\BIBentryALTinterwordspacing
M.~Varasteh, B.~Rassouli, and B.~Clerckx, ``On capacity-achieving distributions
  for complex {AWGN} channels under nonlinear power constraints and their
  applications to {SWIPT},'' \emph{CoRR}, vol. abs/1712.01226, 2017. [Online].
  Available: \url{http://arxiv.org/abs/1712.01226}
\BIBentrySTDinterwordspacing

\bibitem{OShea_Hoydis_2017}
T.~O'Shea and J.~Hoydis, ``An introduction to deep learning for the physical
  layer,'' \emph{IEEE Trans. Cognitive Communications and Networking}, vol.~3,
  no.~4, pp. 563--575, Dec. 2017.

\bibitem{Nachmani_etall}
E.~Nachmani, E.~Marciano, L.~Lugosch, W.~J. Gross, D.~Burshtein, and Y.~Be'ery,
  ``Deep learning methods for improved decoding of linear codes,'' \emph{IEEE
  Journal of Selected Topics in Signal Processing}, vol.~12, no.~1, pp.
  119--131, Feb. 2018.

\bibitem{Caciularu_Burshtein}
A.~Caciularu and D.~Burshtein, ``Blind channel equalization using variational
  autoencoders,'' in \emph{IEEE Int'l Conf. Commun. Workshops (ICC Workshops)},
  May 2018, pp. 1--6.

\bibitem{Felix_Cammerer_Dorner}
A.~Felix, S.~Cammerer, S.~Dorner, J.~Hoydis, and S.~T. Brink,
  ``{OFDM}-{A}utoencoder for end-to-end learning of communications systems,''
  in \emph{IEEE 19th Int'l Workshop on Signal Processing Advances in Wireless
  Communications (SPAWC)}, Jun. 2018, pp. 1--5.

\bibitem{Varasteh_Piovano_Clerckx}
M.~{Varasteh}, E.~{Piovano}, and B.~{Clerckx}, ``A learning approach to
  wireless information and power transfer signal and system design,'' in
  \emph{ICASSP}, May 2019, pp. 4534--4538.

\bibitem{Vedady_Zeng}
M.~R.~V. Moghadam, Y.~Zeng, and R.~Zhang, ``Waveform optimization for
  radio-frequency wireless power transfer : (invited paper),'' in \emph{IEEE
  18th Int'l Workshop on Signal Processing Advances in Wireless Communications
  (SPAWC)}, Jul. 2017, pp. 1--6.

\bibitem{Valenta_Durgin}
C.~R. Valenta and G.~D. Durgin, ``Harvesting wireless power: Survey of
  energy-harvester conversion efficiency in far-field, wireless power transfer
  systems,'' \emph{IEEE Microwave Magazine}, vol.~15, no.~4, pp. 108--120, Jun.
  2014.

\bibitem{Hoydis_Aoudia}
\BIBentryALTinterwordspacing
F.~A. Aoudia and J.~Hoydis, ``Model-free training of end-to-end communication
  systems,'' \emph{CoRR}, vol. abs/1812.05929, 2018. [Online]. Available:
  \url{http://arxiv.org/abs/1812.05929}
\BIBentrySTDinterwordspacing

\end{thebibliography}

\end{document}